\documentclass[twocoumn]{IEEEtran}

\title{Effect of LOS/NLOS Propagation on Area Spectral Efficiency and Energy Efficiency of Small-Cells}
\author{Carlo Galiotto,~~Ismael Gomez-Miguelez,~~Nicola Marchetti,~~Linda Doyle\\
CTVR, Trinity College, Dublin, Ireland \\ \{galiotc,~gomezi,~marchetn,~Linda.Doyle\}@tcd.ie.
}

\usepackage[dvips]{graphicx}
\usepackage[T1]{fontenc}
\usepackage{amsmath}
\usepackage{amssymb}
\usepackage{amsthm}
\usepackage{amssymb}
\usepackage{amsfonts}
\usepackage[nolist]{acronym}
\usepackage{microtype}
\usepackage{tabularx}
\usepackage{multirow}
\usepackage{latexsym}
\usepackage{eurosym}
\usepackage{here}
\usepackage{epsfig}
\usepackage{color}      
\usepackage{algorithmic}
\usepackage{subfigure,stfloats}         
\usepackage{float}                    
\usepackage{lettrine}                

\begin{document}

\bstctlcite{IEEEexample:BSTcontrol}
\maketitle
\begin{abstract}
In this paper we investigate the effect of Line-of-Sight (LOS) and Non-Line-of-Sight (NLOS) propagation on the Area Spectral Efficiency (ASE) and on the energy efficiency of dense small-cell networks.  
We show that including both LOS and NLOS propagation in the path-loss model provides a completely different picture of the behaviours of ASE and energy efficiency than what would be observed in case of either LOS or NLOS propagation only. In particular, with combined LOS/NLOS path-loss, the ASE exhibits superlinear and sublinear behaviour at low and high cell densities, respectively. In addition, the energy efficiency as a function of the cell density has a global maximum and is not a monotonically increasing function like in case of LOS or NLOS propagation only. Based on our findings, we claim that Line-of-Sight (LOS) and Non-Line-of-Sight (NLOS) propagations play an important role in studying the performance of extremely dense small-cell networks.
\end{abstract}



\section{Introduction}

\label{sect:intro}

Motivated by the encouraging predictions on performance enhancement of wireless networks enabled by cell densification, network operators are showing huge interest in small-cells. In fact, in common scenarios cell densification is expected to provide a linear gain in terms of
throughput delivered by the network as the number of nodes increases
\cite{Andrews2011}. In addition, the overall transmit power used
by all the base stations decreases with the density of nodes, while
guaranteeing linear throughput gain \cite{Galiotto2013}. As a consequence of these two results, we will show later in this paper that the energy efficiency under full buffer traffic regime is monotonically increasing, meaning that the efficiency also increases with the node density. 

Nonetheless, as such results seem to be too optimistic, it is reasonable to question whether they are generally true or, instead, if they come as a  consequence of too simplistic system models.
Hence, in this paper we investigate the effect of a combined Line-of-Sight (LOS) and Non Line-of-Sight (NLOS) propagation model on the Area Spectral Efficiency (ASE) gain and we study the related energy efficiency of cell-splitting. We show that modelling the signal propagation according to the combined LOS/NLOS
path-loss provides a different picture of the ASE and of the
energy efficiency behaviour than what would emerge if either LOS or NLOS propagation only were used instead. 

In particular, the main contributions of this paper are the following. First, we show that the ASE gain is no longer a linear function of the cell density but, as a consequence of the combined LOS/NLOS propagation, it exhibits superlinear behaviour at low cell densities and sublinear behaviour at high cell densities. Second, we show that the energy efficiency is not a monotonically increasing function but has a global maximum for a given node density. 

Therefore, on the basis of our findings, we make the argument that the signal propagation model has a strong impact on the performance trend in small-cell networks and that particular attention should be paid in choosing the correct
propagation model when addressing some specific investigations in extremely dense networks.

The remainder of this paper is organized as follows. In Section \ref{sect:RelatedWork}
we discuss the main results of the related work on ASE and energy efficiency
in small-cells networks. In Section \ref{sec:systemModel} we describe the path-loss model and the methodology applied in our study. We then provide the results of the ASE and of the transmit power per base station in Section \ref{sec:Simulation_Results_part_I}. In Section \ref{sec:Energy_Efficiency}
 we discuss how the LOS/NLOS propagation affects the energy efficiency and we present the simulation results. Finally, conclusions
are drawn in Section \ref{sec:conclusions}.

\section{Related work}
\label{sect:RelatedWork}

Investigation on the throughput gain achieved by increasing the cell-density in wireless networks has been done in recent years. The authors in \cite{Andrews2011} have shown analytically that when the network is interference limited and the base stations are distributed according to a Spatial Poisson Point Process (SPPP), increasing the cell density yields linear increase of the network throughput. Regarding the transmit power, an investigation on the overall transmit power needed by the network to achieve linear ASE gain has been proposed in \cite{Galiotto2013}.  
However, a single slope propagation model is assumed for the studies proposed in both \cite{Andrews2011} and \cite{Galiotto2013}. 

Some work on area spectral efficiency gain of cell splitting under
different propagation models can be found in the literature. The authors
in \cite{Ling2012} addressed the problem of assessing the throughput
gain in indoor scenarios and specifically, they considered a propagation model proposed by \cite{Medbo2000}
which has an exponential component. Under this assumption, the area spectral efficiency gain is proved to scale as $\sqrt{N}$, where $N$ is the number of nodes. However, this propagation model has only been proposed for indoor scenarios. Furthermore, the authors of \cite{Ling2012} address neither the transmit power nor the energy efficiency related to this propagation model.

The problem of the energy efficiency vs spectral efficiency trade-off
has been investigated in \cite{Rao2013}. The authors of \cite{Rao2013}
make use of stochastic geometry to carry out an analysis of the optimal
energy efficient and of the optimal spectral efficient regimes of the network.
Nonetheless, the problem formulation proposed in \cite{Rao2013}
assumes that the user is kept at a fixed distance from the base station
and this distance remains unchanged while the node density varies. Although this 
is a reasonable assumption in scenarios where there are small variations
of the node density, in our paper we investigate the effect of large
variations of the cell density. Hence, the model proposed in \cite{Rao2013}
would not be accurate for our analysis. Moreover, a single slope propagation model is assumed by the authors in \cite{Rao2013}.

\section{Models and methodology} \label{sec:systemModel}

\subsection{Propagation model}\label{subsec:propagationModel}

In this paper we consizder  the combined LOS/NLOS path-loss
model recommended by the 3rd Generation Partnership Project (3GPP)
for studying Heterogeneous Networks. In particular, we opted for
the outdoor propagation model for pico-cells/hotzones \cite{3GPP36814}
which is the following: 
\begin{equation}
\mathrm{PL_{\mathrm{L/NL}}}(d)=\begin{cases}
K_{\mathrm{L}}d^{\beta_{\mathrm{L}}} & \text{with probability}\; p_{\mathrm{L}}(d),\\
K_{\mathrm{NL}}d^{\beta_{\mathrm{NL}}} & \text{with probability}\:1-p_{\mathrm{L}}(d),
\end{cases}\label{eq:propag_outdoor}
\end{equation}
where $\beta_{\mathrm{L}}$ and $\beta_{\mathrm{NL}}$ are the path-loss exponents for LOS and NLOS propagation, respectively; $K_{\mathrm{L}}$ and $K_{\mathrm{NL}}$ are the signal attenuations at distance $d=1$  for LOS and NLOS propagation, respectively; $p_{\mathrm{L}}(d)$ is the probability that the base station is in line-of-sight with the user; $p_{\mathrm{L}}(d)$ is a function dependent on the base station-to-user distance $d$  and is given by the following equation: 
\begin{equation}
p_{\mathrm{L}}(d)=0.5-\min(0.5,5\,\mathrm{e}^{\frac{-d_0}{d}})+\min(0.5,5\,\mathrm{e}^{\frac{-d}{d_1}}).\label{eq:probability_LOS}
\end{equation}
In this paper we refer to the path loss model described by eq. \eqref{eq:propag_outdoor}
and \eqref{eq:probability_LOS} as \textit{combined LOS/NLOS} propagation
model. On the contrary, we refer to the more common model $\mathrm{PL}_{\mathrm{SL}}(d)=K_{\mathrm{SL}}d^{\beta}$
as \textit{single slope} propagation model, as in a logarithmic scale
it becomes a linear function. 

With the values of the parameters $K_{\mathrm{L}}$, $\beta_{\mathrm{L}}$, $K_{\mathrm{NL}}$, $\beta_{\mathrm{NL}}$, $d_0$ and $d_1$ which will be given later in Table \ref{table:simulation-parameters}, the NLOS channel attenuation  increases with a higher slope than the LOS one. Also, the probability of having LOS propagation decreases with the distance. Overall, with the combined LOS/NLOS model, the ASE and the energy efficiency  in small-cells networks show different behaviours than those observed with the single slope model. 

\vspace{-3mm}
\subsection{Methodology for analysis} \label{sec:Methodology}

In this paper we first make use of simulation results to assess the ASE and the transmit power in dense networks with combined LOS/NLOS propagation model. We then apply curve fitting for extrapolating the trend of the ASE and of the transmit power as functions of the node density. Finally, by means of the mathematical functions obtained through data fitting, we analyze the behaviour of the energy efficiency and we explain why the LOS/NLOS propagation model yields different results on the energy efficiency as a function of the node
density compared to the single slope propagation model. 

\noindent We obtained the results following the steps reported below: 
\begin{enumerate}
\item[i.] We use a system level simulator to compute the wideband  Signal to Interference Ratio (SIR)
of the users. We first assume the network is interference limited and
that all the base stations transmit over the same band, i.e., reuse
1 is used. We run simulations for different values of cell density. 
\item[ii.] We compute the transmit power necessary in order to ensure the users
experience coverage and throughput as close as possible to the case
of interference limited network as the cell density varies. Our criterion
to compute the transmit power will be explained later in Section \textbf{\eqref{sub:Computing-the-transmit}}. We then compute the ASE.
\item[iii.] We use curve fitting to determine how the ASE and the transmit
power scale with the node density. Based on these throughput and power trends
we also compute the energy efficiency (see Section \ref{sec:Energy_Efficiency}). 
\end{enumerate}

\subsubsection{Computing the transmit power per base station}\label{sub:Computing-the-transmit}

The transmit power per base station should be set in order to guarantee that the network remains in interference-limited regime, i.e.,  the transmit power should be high enough so that the thermal noise power at the user receiver can be neglected with respect to the interference power at the receiver. Under this condition, the SINR will not be limited by the transmit power. In fact, with the network in interference limited regime, the transmit power is high enough that any further increase of it would be pointless in terms of enhancing the SINR, since the receive power increment would be balanced by the exact same interference increment. 

If we translate this concept into the analysis of SINR Cumulative Distribution Function (CDF), we have that the SIR CDF $F_{\xi}(y)=\mathrm{P}[\xi\leq y]$ is the limit of the SINR CDF $F_{\gamma}(y)=\mathrm{P}[\gamma\leq y]$ as the transmit power tends to infinity. Hence, to keep the network in interference-limited regime, the power should be set to a value high enough to guarantee the SINR CDF curve to be close to its upper bound, i.e,  the SIR CDF curve. To achieve this, we impose a condition on the difference between the SIR
and SINR CDFs i.e., we compute the minimum transmit power per base station so that the inequality $\Delta\mathrm{dB}(Y_{\%})=F_{\xi}^{-1}(Y_{\%})-F_{\gamma}^{-1}(Y_{\%})\leq\Delta\mathrm{dB}_0$ is verified.

For setting the numerical values of $Y_{\%}$ and $\Delta\mathrm{dB}_0$ we consider the users located close to the serving base station which usually experience low interference and high SINR. The SINR of these users is sensitive to the changes of the transmit power, as the interference here is less severe and then the user might not be in interference-limited regime. This means that imposing $\Delta\mathrm{dB}(Y_{\%}) \leq \Delta\mathrm{dB}_0$ for high values of SINR (equivalently, for high percentages $Y_{\%}$) is a stricter condition on the power than for low values of SINR. We believe that the values $Y_{\%}=80\%$ and $\Delta\mathrm{dB}_0=0.2\mathrm{dB}$ we set in our simulation are strict enough to guarantee the network to be interference-limited.

\subsubsection{Curve fitting}

We use a function of the kind $f(z)=az^{b}$ to interpolate the simulation results of ASE and transmit power per base station. We chose this function because it is suitable to fit sets of ASE data with linear behaviour (i.e., $b=1$), superlinear behaviour (i.e., $b>1$) and sublinear beviour (i.e., $b<1$). Moreover, as in logarithmic scale the function $f(z)=az^{b}$ turns to be a linear function whose slope (in logarithmic scale) is $b$, we are particularly interested in determining the values of $b$, as this parameter characterizes the steepness
of the ASE and of the transmit power as a function of the node density. Also the function $f(z)=az^{b}$ turns to be mathematically tractable when,
later on in Section \eqref{sub:Energy-efficiency-for}, we need to
compute the derivative of the energy efficiency.

\section{Simulation results on ASE and on transmit power} \label{sec:Simulation_Results_part_I}

In this Section we present the simulation results we obtained with
the parameters setting given in Table \ref{table:simulation-parameters}.
Regarding the scenario, we consider a small-cell network where the
base stations distribution follows both a regular geometric
pattern and a Spatial Poisson Point Process (SPPP) model. We assume the
base stations to be pico base stations with omni-directional antennas. Nonetheless, since the two models provide similar results, we will show the results regarding the SPPP only for the ASE; we will then consider the square grid model for the transmit power per base station and for the energy efficiency later in Section \ref{sec:Energy_Efficiency}.

\begin{table}[tbph]
\caption{Simulation parameters}
\label{table:simulation-parameters}
\centering%
\begin{tabular}{|p{2.55cm}|p{5.5cm}|}
\hline 
\textbf{Parameter}  & \textbf{Value}\tabularnewline
\hline 
\hline 
\multirow{2}{*}{Scenarios}  & i) Base stations placed in a $1000\,\mathrm{m}\times1000\,\mathrm{m}$
square grid.\tabularnewline
\hline 
 & ii) SPPP over a $1000\,\mathrm{m}\times1000\,\mathrm{m}$ square\tabularnewline
\hline 
User distribution  & Uniform distribution \tabularnewline
\hline 
Number of snapshots  & 50\tabularnewline
\hline 
Path-loss - Single slope & \multirow{1}{5.5cm}{$\mathrm{PL}_{\mathrm{SL}}(d_{\mathrm{km}})=140.7+36.7\log(d_{\mathrm{km}})$,~~$\beta=3.67$, ~~$K_{\mathrm{SL}}=10^{14.07}$  \cite{3GPP36814}}\tabularnewline
\hline 
Path-loss - Combined LOS/NLOS & See \eqref{eq:propag_outdoor}; with $d$ in km, $K_{\mathrm{L}}=10^{10.38}$, $\beta_{\mathrm{L}}=2.09$,
$K_{\mathrm{NL}}=10^{14.54}$, $\beta_{\mathrm{NL}}=3.75$, $d_0 =0.156\mathrm{km}$, $d_1 = 0.03\mathrm{km}$ \;\cite{3GPP36814}\tabularnewline
\hline 
Shadow fading  & Log-normal, 8 dB standard deviation \cite{3GPP36814}\tabularnewline
\hline 
Penetration loss  & 20 dB \cite{3GPP36814}\tabularnewline
\hline 
Bandwidth $\mathrm{BW}$ & 10 MHz centered at 2 GHz. All the base stations transmit over the
same band, i.e., reuse 1 is used.\tabularnewline
\hline 
Noise  & Additive White Gaussian Noise with -174 dBm/Hz Power Spectral Density\tabularnewline
\hline 
Noise Figure  & 9 dB\tabularnewline
\hline 
Capacity function  & $c(\gamma)=\max\left(0.75\log_{2}\left(1+\frac{\gamma}{1.33}\right),5.55\right)$
\cite{Harri2009}\tabularnewline
\hline 
Antenna at BS and UE  & Omni-directional with 0 dBi gain\tabularnewline
\hline 
$K_{\mathrm{RF}}$ & $10$ \cite{Auer2011}\tabularnewline
\hline 
$P_{0}$ & $2$ W, $10$ W \cite{Auer2011}\tabularnewline
\hline 
\end{tabular}

\end{table}


\vspace{-5mm}
\subsection{Area spectral efficiency} \label{sub:Area_Spectral_Efficiency}

\begin{figure*}[]
\centering
    \subfigure[\label{fig:ASE_linear}]{
\includegraphics[width=.92\columnwidth]{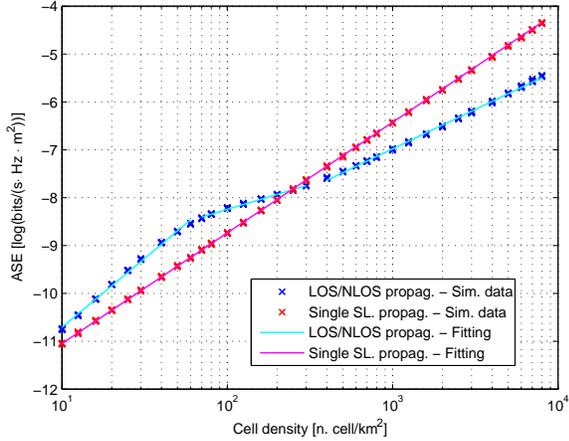}}
\hspace{1em}
    \subfigure[\label{fig:ASE_log}]{
\includegraphics[width=.92\columnwidth]{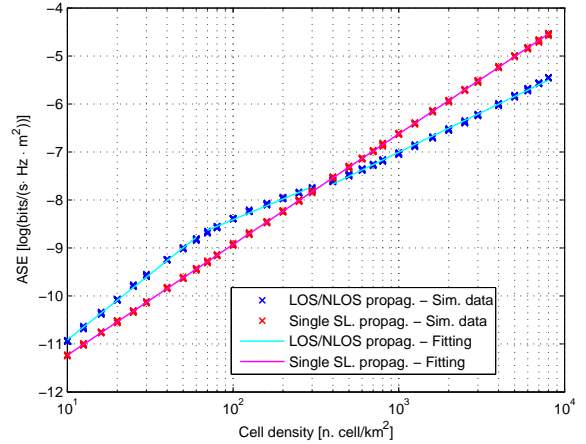}
\vspace{-7mm}
}
\caption{ASE vs cell density: square grid (a) and SPPP based model (b).}
\label{fig:ASE}
\vspace{-6mm}
\end{figure*}

We define the area spectral efficiency as the network throughput normalized with respect to the bandwidth and the area. We first compute the SINR $\gamma$ and then the spectral efficiency is obtained as a
function $c(\gamma)$ of $\gamma$ (see Table \ref{table:simulation-parameters});
the ASE is then calculated by summing up the average cell spectral efficiencies of all the cells in the network  and then by normalizing with respect to the area. The plots in Fig. \ref{fig:ASE} show the ASE simulation results and the corresponding fitting curves obtained through linear regression with least square solution for \textit{single slope} and \textit{combined LOS/NLOS} path-loss models. Fig. \ref{fig:ASE_linear} shows the results obtained with the square grid model whereas Fig. \ref{fig:ASE_log} shows the results obtained with SPPP model.

If we look at the plot in Fig. \ref{fig:ASE_linear} and \ref{fig:ASE_log}, there is a noticeable  difference  between the two cases of \textit{single slope} and \textit{combined LOS/NLOS} propagation models; while with single slope path-loss the ASE grows linearly with the density of nodes, with combined LOS/NLOS path-loss, the curve we obtain does not show linear behaviour. 
Moreover, in order to improve the accuracy of the fitting for the combined LOS/NLOS path-loss, the data interpolation has been done through a piece-wise linear multi-slope function (in the logarithmic domain).
This piece-wise function consists of three functions in the form of
$\eta_{\mathrm{A}}(x)=\eta_{\mathrm{A},0}x^{\alpha}$ where each of these functions
is specified within a finite range of node density values. The parameters
$\eta_{\mathrm{A},0}$ and $\alpha$ which correspond to the different
intervals of density values $x$ are reported in Table \ref{table:ASEresults}; to obtain these values we assume the ASE to be in bits/(s$\cdot$ Hz $\cdot$ m$^2$) and the density $x$ to be in number of BSs per m$^2$.

\begin{table}[tbph]
\caption{ASE fitting curves parameters}
\label{table:ASEresults}
\centering

\begin{tabular}{|c|c|c|c|c|}
\hline 
\multirow{2}{*}{$x$ intervals [n. cell/km$^{2}$]} & \multicolumn{2}{c|}{Square grid.} & \multicolumn{2}{c|}{SPPP}\tabularnewline
\cline{2-5} 
 &  $\eta_{\mathrm{A},0}$ & $\alpha$  &$\eta_{\mathrm{A},0}$   & $\alpha$  \tabularnewline
\hline 
 $\mathcal{D}_{\mathrm{1}}:\,x\in[10,\,60)$ &  $3.98\cdot10^{1}$ &  $1.25$& $1.65\cdot10^{1}$ & $1.19$ \tabularnewline
\hline 
 $\mathcal{D}_{\mathrm{2}}:\,x\in[60,\,400)$ & $1.64\cdot10^{-2}$ &  $0.45$ & $6.79\cdot10^{-2}$  & $0.62$ \tabularnewline
\hline 
 $\mathcal{D}_{\mathrm{3}}:\,x\in[400,\,8000)$ & $1.30\cdot10^{-1}$ & $0.72$ & $1.34\cdot10^{-1}$ & $0.72$  \tabularnewline
\hline 
\end{tabular}

\end{table}

In case of single slope propagation, the ASE
scales nearly linearly with the node density. In fact, the exponent $\alpha$
of the interpolating function we obtain by fitting the data is 1.0019,
meaning that $\eta_{\mathrm{A}}(x)=\eta_{\mathrm{A},0}x^{\alpha}$ is nearly linear.

On the other hand, in case of combined LOS/NLOS path-loss, the slope of the ASE as a function of the density $x$ changes depending on $x$ itself.
At low densities, the base stations in the network are sparse and then, on average, users are located far away from the BSs. For this reason, as the probability of having LOS decreases with distance, at low density (e.g., for $x$ close to $10$ BSs/km$^2$) both the serving base station and the interferers are likely to have NLOS with the user. However, with reference to Table  \ref{table:ASEresults}, when the density increases within range $\mathcal{D}_{\mathrm{1}}$ the probability of having the serving BS in LOS with the user increases much faster then the probability of having the interferer in LOS with the same user; as a consequence, the attenuation of the received signal decreases much faster than the attenuation of the interfering signal, leading to a considerable SINR  gain. Hence, within range $\mathcal{D}_{\mathrm{1}}$, increasing the density results in a ASE gain $\alpha>1$ which is higher than the linear gain $\alpha=1$. 

Nonetheless, as the density keeps increasing, the probability that some interferers enter the LOS region increases, making the interference to the users stronger. This explains why within the range $\mathcal{D}_{\mathrm{2}}$ the slope of the ASE drops down to $\alpha=0.45$ and $\alpha=0.62$ for the square grid model and for the SPPP model, respectively. Finally, once most of the strongest interferers will have entered the LOS region, further increases of density will result in a slightly higher ASE gain, which settles to a value of $\alpha=0.72$ for density within the range $\mathcal{D}_{\mathrm{3}}$. Let us notice that, although the value $\alpha=0.72$ is bigger than $\alpha=0.45$ (and than $\alpha=0.62$) as the most of the interfering BSs have entered the LOS region, there are still new interferers entering the LOS region as the density increases within the range $\mathcal{D}_{\mathrm{3}}$; this explain why the ASE gain $\alpha=0.72$ is sublinear.

Hence, the ASE prediction attained by using single slope propagation
model represents an optimistic estimation of what would be observed
instead with combined LOS/NLOS propagation as the density increases. Therefore, we may infer that, in extremely dense networks, the combined LOS/NLOS path-loss should be preferred to the single slope propagation model.

\subsection{Transmit power per base station}\label{sub:TXPowerPerBS}

\begin{figure}
\centering
\includegraphics[width=.92\columnwidth]{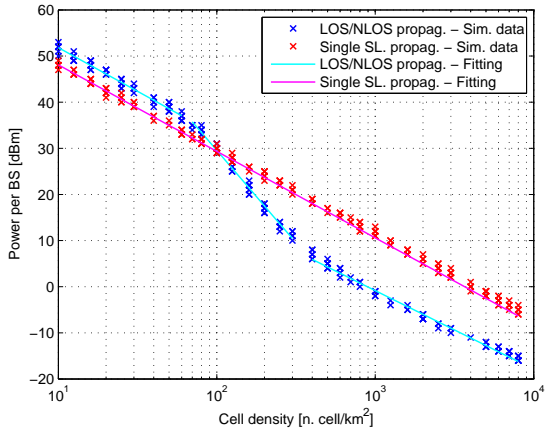}
\caption{Transmit power per BS - square grid model.}\label{fig:transmit_power}
\vspace{-4mm}
\end{figure}

In Fig. \ref{fig:transmit_power} we show the simulation results of
the transmit power per base station $P_{\mathrm{TX}}(x)$, which has been computed as explained in Section
\ref{sub:Computing-the-transmit}. In this figure we compare the results
we obtained using the \textit{single slope} and the \textit{combined LOS/NLOS}
path loss models; the lines superimposed on the dots represent the
curves used for fitting which have been obtained by means
of linear regression with least squares solution.

As we already observed for the ASE, the functions describing how $P_{\mathrm{TX}}(x)$ decays with the BS density $x$ is different in the two cases of single slope and combined LOS/NLOS propagation. With reference to Fig. \ref{fig:transmit_power}, in case of single slope path loss the power decreases linearly (in logarithmic scale) with the density and 
the slope of the
line used to fit the transmit power in case of the single slope propagation
model is -1.83 which, as anticipated in \cite{Galiotto2013}, is
close to the value $-\beta/2$, with $\beta$ being the exponent
of the single slope propagation model (see Table \eqref{table:simulation-parameters}).
In the case of combined LOS/NLOS propagation, the data interpolation
has been done through a piece-wise linear multi-slope function (in
the logarithmic domain) in order to improve the accuracy of the fitting.
The parameters
$P_{\mathrm{T}}$ and $\delta$ of the fitting functions $P_{\mathrm{TX}}(x)=P_{\mathrm{T}}x^{\delta}$ which correspond to the different
intervals of density values $x$ are reported in Table \ref{table:POWERresults}; to obtain these fitting values we assume  $P_{\mathrm{TX}}$ to be in Watts and the density $x$ to be in number of BSs per m$^2$.

\begin{table}[tbph]
\caption{Transmit power fitting curves parameters}
\label{table:POWERresults}
\centering
\begin{tabular}{|c|c|c|}
\hline 
Density interval [n. cell/km$^{2}$] &  $P_{\mathrm{T}}$ & $\delta$ \tabularnewline
\hline 
\hline 
$\mathcal{D}_{\mathrm{1}}:\,x\in[10,\,60)$ &  $4.516\cdot10^{-8}$ & $-1.90$ \tabularnewline
\hline 
$\mathcal{D}_{\mathrm{2}}:\,x\in[60,\,400)$ &  $7.210\cdot10^{-17}$ & $-4.01$ \tabularnewline
\hline 
$\mathcal{D}_{\mathrm{3}}:\,x\in[400,\,8000)$ &  $5.949\cdot10^{-9}$ & $-1.70$ \tabularnewline
\hline 
\end{tabular}

\end{table}
The fact that the transmit power per base station decays more or less steeply with the density $x$ depends on how quickly the interference power increases or decreases with $x$. As we explained in Section \ref{sub:Computing-the-transmit}, the transmit power per base station $P_{\mathrm{TX}}(x)$ has to be set so that the network is interference limited. Thus, if the channel attenuation between the interferer and the user decreases quickly as the density increases, a lower transmit power will be enough to guarantee that the interference power is greater than the noise power.
In other words, if the interferer-to-user channel attenuation tends to decrease quickly as the density increases, so does the transmit power and vice-versa.

This explains why within range  $\mathcal{D}_{\mathrm{2}}$ the transmit power $P_{\mathrm{TX}}(x)$ decreases steeply with slope $\delta=-4.01$. In fact, as explained in Section \ref{sub:Area_Spectral_Efficiency}, as $x$ increases within range  $\mathcal{D}_{\mathrm{2}}$, the probability of having interferers in LOS with the user rises and, as a consequence, we have a lower attenuation of the channel between the interfering base station and the user.
Hence, the $P_{\mathrm{TX}}(x)$ which guarantees the network to be in the interference-limited regime will also decrease steeply as $x$ increases.
On the contrary, within range $\mathcal{D}_{\mathrm{3}}$, most of the interferers will have already entered the LOS zone (see Section \ref{sub:Area_Spectral_Efficiency}), meaning that the interferer-to-user channel attenuation drops less rapidly than within range $\mathcal{D}_{\mathrm{2}}$; for this reason, also $P_{\mathrm{TX}}(x)$ will decrease less rapidly than within range $\mathcal{D}_{\mathrm{2}}$.

Let us note that the transmit power per base station as indicated
in Fig. \ref{fig:transmit_power} tends to be low, as it goes
even below 0 dBm for densities higher than $10^{3}$cells/km$^{2}$.
If we compare these values with the indications given by 3GPP documents
\cite{3GPP36814} for simulations of pico-cell scenarios (suggested
power per pico cells is between 30dBm and 20dBm), we notice a substantial
difference. However, the 3GPP recommendations for the pico-base station
transmit power are meant for isolated cells covering hotzones; on
the contrary, in our scenario the network is entirely covered by small
cells. As a cell in a dense network  covers a considerably smaller
area than in the case of isolated cells, the required transmit power
per base station is much lower compared to the case of an isolated
cell.

\section{Energy efficiency} \label{sec:Energy_Efficiency}

Regarding the computation of the energy consumption we assume a fully loaded network, i.e., there are more users than BSs and every BS transmits to a non-empty set of users. We further assume full buffer traffic and full spectrum reuse; hence, there are always data to transmit and the BSs use all the time and frequency resources available.  

We model the power consumption $P_{\mathrm{BS}}$ of the base station
assuming that $P_{\mathrm{BS}}$ is the sum of two components, i.e.,
$P_{\mathrm{BS}}=P_{0}+P_{\mathrm{RF}}$; the first, denoted by $P_{0}$,
takes into account the energy necessary for signal processing and
to power up the base station circuitry. This power $P_{0}$ is modelled
as a component being independent of the transmit power and of the
base station load \cite{Auer2011}. The second component, denoted
by $P_{\mathrm{RF}}$, takes into account the power fed into the power
amplifier which is then radiated for signal transmission. The power
$P_{\mathrm{RF}}$ is considered to be proportional to the power transmitted
by the base station; we can thus write  $P_{\mathrm{RF}}=K_{\mathrm{RF}}P_{\mathrm{TX}}$,
where $K_{\mathrm{RF}}$ takes into account the losses of the power
amplifier (i.e., we assume $K_{\mathrm{RF}}$ to be the inverse of the power amplifier efficiency). 

Under these assumptions, the power consumed by the cellular network
made of $N$ base stations can be written as follows: 
\begin{equation}
P_{\mathrm{TOT}}=NP_{\mathrm{BS}}=NP_{0}+NK_{\mathrm{RF}}P_{\mathrm{TX}}.
\end{equation}
where $N$ can also be written as $N=Ax$, with $A$ denoting the area
of the network and $x$ denoting the density of base stations. As we
have shown in Section \ref{sub:TXPowerPerBS}, the transmit power
$P_{\mathrm{TX}}$ varies as a function of $N$ or, equivalently, of $x$.

In this paper we are interested in characterizing the energy efficiency
of the network as a function of the node density to identify
the trade-off between the area spectral efficiency and the power consumed
by network. We define the \textit{energy efficiency} as the ratio
between the overall throughput delivered by the network and the total
power consumed by the wireless network. We can write the energy efficiency
as follows: 
\begin{equation}
\eta_{\mathrm{EE}}(x)\triangleq\frac{R(x)}{P_{\mathrm{TOT}}(x)},\label{eq:eff_definition}
\end{equation}
where $R(x)$ is the network throughput which can be written as $R(x)= A\cdot\mathrm{BW}\cdot \eta_{\mathrm{A}}(x)$,
with \textbf{$\mathrm{BW}$} denoting the bandwidth and $\eta_{\mathrm{A}}(x)$ denoting the area spectral efficiency.

\subsection{Energy efficiency for single slope propagation} \label{sub:Energy_efficiency_for_LOS}

When the path-loss can be expressed as $\mathrm{PL}_{\mathrm{SL}}(d)=K_{\mathrm{SL}}d^{\beta}$,
increasing the node density yields linear throughput gain \cite{Galiotto2013}, \cite{Ling2012}; therefore, the throughput can be written in the form $R(x)=AR_{0}x$, where $R_0$ is a constant. Also, the transmit power of each base station
can be scaled as $P_{\mathrm{TX}}(x)=P_{1}x^{-\beta/2}$ \cite{Galiotto2013}; the effect of the power amplifier losses is taken into account in $P_1$. Hence, the energy efficiency becomes: 
\begin{equation}
\eta_{\mathrm{EE}}(x)=\frac{R_{0}x}{xP_{0}+xP_{1}x^{-\beta/2}}=\frac{R_{0}}{P_{0}+P_{1}x^{-\beta/2}}.\label{eq:eff_pathloss_1}
\end{equation}
If we compute the derivative of $\eta_{\mathrm{EE}}(x)$, we obtain
$\frac{\mathrm{d}\eta_{EE}(x)}{\mathrm{d}x}=\frac{R_{0}P_{1}\frac{\beta}{2}x^{-\beta/2-1}}{\left(P_{0}+P_{1}x^{-\beta/2}\right)^{2}}$,
which is strictly positive for any node density value $x>0$. Hence,
under the assumption of single slope propagation model, increasing
the node density yields higher energy efficiency independently of the value of $\beta>0$ of the propagation model. Nonetheless, as the density keeps increasing, the transmit power will become negligible compared to $P_0$, meaning that the power consumption will be almost entirely impacted by the power offset $P_0$. Hence, we obtain that $\lim_{x \to \infty} \eta_{\mathrm{EE}}(x) = R_0/P_0$, i.e., the energy efficiency saturates and converges to $R_0/P_0$.

\vspace{-6mm}
\subsection{Energy efficiency for combined LOS/NLOS propagation}\label{sub:Energy-efficiency-for}

As we mentioned in Section \ref{sub:Computing-the-transmit}, we will use functions of the
kind $f(z)=az^{b}$ to interpolate both the data on area spectral
efficiency (and then throughput) and also on the transmit power per
base station. We then expect the throughput to be in the form of $R(x)=AR_{1}x^{\alpha}$
and the power in the form $P_{\mathrm{TX}}=P_{\mathrm{T}}x^{\delta}$.
Thus, with this assumption, we obtain the following expression for
the energy efficiency 
\begin{equation}
\eta_{\mathrm{EE}}(x)=\frac{R_{1}x^{\alpha}}{xP_{0}+xP_{\mathrm{T}}x^{\delta}}=\frac{R_{1}x^{\alpha-1}}{P_{0}+P_{\mathrm{T}}x^{\delta}}.\label{eq:eff_pathloss_2}
\end{equation}
The derivative of $\eta_{\mathrm{EE}}(x)$ is given below: 
$$
\frac{\mathrm{d}\eta_{EE}(x)}{\mathrm{d}x}=\frac{R_{1}P_{0}(\alpha-1)x^{\alpha-2}+R_{1}P_{\mathrm{T}}(\alpha-\delta-1)x^{\alpha+\delta-2}}{\left(P_{0}+P_{\mathrm{T}}x^{\delta}\right)^{2}}.
$$
Depending on the value of $\alpha$ and of $\delta$, there can exist optimum points of the function $\eta_{\mathrm{EE}}(x)$. In the following paragraphs, we analyze the derivative $\eta\prime_{\mathrm{EE}}(x)=\frac{\mathrm{d}\eta_{EE}(x)}{\mathrm{d}x}$ in order to assess the existence of such optima. Let us note that $R_{1}$, $P_{0}$ and $P_T$ are positive; moreover it is reasonable to assume that $\alpha>0$ (i.e., the area spectral efficiency is an increasing function of the density) and that $\delta<0$, i.e., the transmit power per BS is a decreasing function of the density.
\subsubsection{The energy efficiency is a monotonically increasing function}	If $\alpha>1$, i.e., if the ASE growth is superlinear, then also $\alpha>1>1+\delta$ holds true; in this case, $\eta\prime_{\mathrm{EE}}(x)$ is strictly positive, meaning that energy efficiency increases as the density increases. This can be explained by the fact that the ASE and so the throughput grow faster than the total power used by the network, which implies adding more base stations with lower transmit power improves the energy efficiency.
\subsubsection{The energy efficiency is a monotonically decreasing function}	If $\alpha<1$ (i.e., the ASE growth is sublinear) and $\alpha<1+\delta$, then the $\eta\prime_{\mathrm{EE}}(x)$ is strictly negative and then the energy efficiency is a monotonically decreasing function of the density $x$. This is due to the ASE which grows too slowly compared to the total power consumption of the network, making the addition of base stations in the network inconvenient from the energy efficiency point of view.
\subsubsection{The energy efficiency exhibits an optimum point}	
If $\alpha<1$ (i.e., ASE gain is sublinear) and if $\alpha>1+\delta$, then we obtain that the derivative $\eta\prime_{\mathrm{EE}}(x)$
nulls for $x_{0}=\left(\frac{P_{0}\left(1-\alpha\right)}{P_{\mathrm{T}}\left(\alpha-\delta-1\right)}\right)^{1/\delta}$, is positive for $x<x_0$ and is negative for $x>x_0$. Therefore, $x_0$ is a global maximum of the energy efficiency. 

This behavior of the spectral efficiency is due to the growth of the ASE which is not fast enough to allow the energy efficiency to be monotonically increasing, but it is not even too slow to observe a continuous drop of the energy efficiency; on the one hand, for low base station densities  (i.e., for $x<x_0$) the ASE gain is high enough to counterbalance the total power increase of the network, making the addition of base stations profitable in terms of energy efficiency. On the other hand, as the base station density reaches $x_0$, the ASE gain is not sufficient to compensate the power consumption increment given by any further addition of base stations in the network.
\textbf{REMARKS}: Equations \eqref{eq:eff_pathloss_1} and \eqref{eq:eff_pathloss_2}
hold true only in case of full loaded networks with full-buffer traffic
model. If other models of traffic or load were used, the energy efficiency
would be different than what given in \eqref{eq:eff_pathloss_1} and
\eqref{eq:eff_pathloss_2}. Nonetheless, studying the energy efficiency
of the network under different traffic models is not within the scope
of this paper.

\vspace{-3mm}
\subsection{Simulation results} \label{sub:Energy-efficiency_results}

\begin{figure}
\centering 
\includegraphics[width=.9\columnwidth]{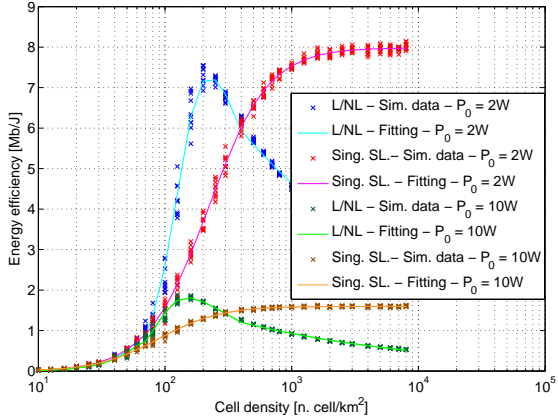}
\caption{Energy efficiency}
\label{fig:energy_efficiency}
\vspace{-5mm}
\end{figure}

In Fig. \ref{fig:energy_efficiency} we show the energy efficiency
results we obtained using single slope and combined LOS/NLOS propagation
model. We considered two different values of the power offset $P_0$, i.e., 2W and 10W. 
As mentioned above in Section \ref{sub:Energy_efficiency_for_LOS},
with single slope propagation the energy efficiency is monotonically
increasing with the node density, meaning the addition of low power
base stations in the network would provide a reward in terms of energy
efficiency. However, as we pointed out in Section \ref{sub:Energy_efficiency_for_LOS}, we can notice that the energy efficiency eventually saturates as the density keeps increasing. 

On the other hand, if we assume combined LOS/NLOS propagation, the
behaviour of the energy efficiency as a function of the node density
looks different. If we compare the values of $\alpha$ and $\delta$ in Table \ref{table:ASEresults} and \ref{table:POWERresults} with the cases we discussed in Section \ref{sub:Energy-efficiency-for}, for low BS densities , i.e., for $x\in\mathcal{D}_{\mathrm{1}}$, the spectral efficiency is an increasing function of $x$. In fact, within this range of densities, the ASE grows quickly meaning that adding base stations with lower power is beneficial in terms of energy efficiency. 

Nonetheless, with the value of $\alpha$ and $\delta$ for densities $x\in\mathcal{D}_{\mathrm{2}}$, the energy efficiency exhibits a maximum which is achieved for $x=180$ and $x=280$ cells/km$^{2}$ for $P_0=10W$ and $P_0=2W$, respectively; beyond these points, the ASE gain is too low to compensate power consumption increase of the denser network, leading to a drop in terms of energy efficiency. 
Finally, with the value of $\alpha$ and $\delta$ for densities $x\in\mathcal{D}_{\mathrm{3}}$, the energy efficiency is still a non-monotonic function with a stationary point; however, this point occurs at $x_1<400$ 	cells/km$^{2}$ and is then outside the range of values $x\in\mathcal{D}_{\mathrm{3}}$ for which the parameters $\alpha$ and $\delta$ are valid. Hence, the energy efficiency is a decreasing function within $\mathcal{D}_{\mathrm{3}}$, since the ASE gain is too low to pay off the network power consumption.

\section{Conclusions} \label{sec:conclusions}

In this paper we have studied the effect of the combined LOS/NLOS
propagation on the area spectral efficiency and energy efficiency
of dense networks. Based on our study, the estimations of the area
spectral efficiency and of  the energy efficiency carried out using a single slope path
loss seem to provide an optimistic prediction of such metrics. In
fact, if we use a combined LOS/NLOS model as the ones recommended
by the 3GPP for simulation of Heterogeneous Networks, the area spectral
efficiency and the energy efficiency as functions of the node density
exhibit different behaviours. In particular, unlike in the case of
single slope propagation model where the energy efficiency is monotonically
increasing, with the combined LOS/NLOS model there exists a maximum
of the energy efficiency.

\end{document}